\documentclass[aps,prb,twocolumn,superscriptaddress,showpacs,showkeys]{revtex4}
\usepackage{amssymb, amsmath, graphicx}
\usepackage{color}

\definecolor{Red}{rgb}{1,0,0}

\begin{document}


\title{Tunneling spectroscopy of $\mathbf{Tl_2Ba_2CaCu_2O_8}$ single crystals \\and thin films}


\author{Daniel Mazur}
\email[Corresponding author: ]{mazudan@iit.edu}
\affiliation{Physics Division, Illinois Institute of Technology, 3101 South Dearborn St., Chicago, Illinois 60616, U.S.A.}
\affiliation{Argonne National Laboratory, Materials Science Division, 9700 S. Cass Ave, Argonne, Illinois 60439, U.S.A.}

\author{J.~F. Zasadzinski}
\affiliation{Physics Division, Illinois Institute of Technology, 3101 South Dearborn St., Chicago, Illinois 60616, U.S.A.}

\author{K.~E. Gray}
\affiliation{Argonne National Laboratory, Materials Science Division, 9700 S. Cass Ave, Argonne, Illinois 60439, U.S.A.}

\author{Y.~C. Ma}
\author{N.~L. Wang}
\affiliation{Beijing National Laboratory for Condensed Matter Physics, Institute of Physics, Chinese Academy of Sciences, Beijing 100080, People's Republic of China}

\author{S.~L. Yan}
\affiliation{Department of Electronics, NanKai University, Tianjin 300071, People's Republic of China}


\date{\today}

\begin{abstract}
Early electron tunneling experiments with the~$\mathrm{Tl_2Ba_2CaCu_2O_8}$ measured $\Delta= 16-28$~meV, while bulk optical experiments measured $\Delta=43-45$~meV. We report new tunneling measurements of $\mathrm{Tl_2Ba_2CaCu_2O_8}$ single crystals and $c$-axis oriented thin films, where the measured $\Delta$ values on both sample types cover the range of $24 - 50$ meV, with more than 70 \% of the junctions displaying $\Delta > 30$~meV. This work resolves the inconsistency between previously published results of tunneling and optical experiments. New tunneling conductances reveal improved superconducting gap region characteristics consistent with $d$-wave symmetry. The tunneling spectra also display features similar to the dip feature observed in $\mathrm{Bi_2Sr_2CaCu_2O_8}$. We discuss our results in context of experiments reported previously on related materials.
\end{abstract}


\maketitle

\section{Introduction}\label{sec:intro}

Superconducting thallium-based cuprates were first made and studied by Sheng \emph{et al.}\cite{Sheng88,Sheng88a,Sheng88b} and {Hazen} \emph{et al.} \cite{Hazen88}, followed by {Ganguli} \emph{et al.} \cite{Ganguli88a}, Parkin \emph{et al.}\cite{Parkin88}, {Liang} \emph{et al.} \cite{Liang88}, and {Krantz} \emph{et al.} \cite{Krantz88} shortly after the discovery of high-temperature superconductivity. The microscopic mechanism of superconductivity in cuprates is not agreed upon, but the macroscopic $d$-wave model\cite{Anderson61,Won94} has been the most successful phenomenological one. While symmetry of the superconducting cuprates is different from conventional superconductors, still the maximal gap size $\mathrm{\Delta_{max}}$ remains a key parameter of superconducting phenomena in cuprates. Continuing discussions are fueled by factors such as the quantitative differences between data measured on overdoped\cite{Ozyuzer00} and underdoped\cite{Miyakawa99} cuprates, or by inconsistencies among superconducting parameters derived from data obtained by different techniques on nominally identical materials\cite{Huang89,Wang03}. Our research presented in this article contributes to the resolution of inconsistency in superconducting gap parameters published to date on the $\mathrm{Tl_2Ba_2CaCu_2O_8}$ cuprate ({Tl-2212}).

Electron tunneling techniques have proven to be important probes of electronic structure of superconductors. Most importantly, they provide a direct measure of the superconducting gap parameter, $\Delta$. Early electron tunneling studies by {Takeuchi} \emph{et al.}\cite{Takeuchi89} and {Huang} \emph{et~al.}\cite{Huang89} measured the {Tl-2212} polycrystals and single crystals. Published values of $\mathrm{\Delta}$ were $\sim$ 25 {meV} and \mbox{16-28 meV}, respectively. In the latter case, a~majority of junctions exhibited $\Delta\sim$ 20 {meV}, significantly smaller than expected based on the $T_c$. For example, optimally doped $\mathrm{Bi_2Sr_2CaCu_2O_8}$ (Bi-2212) consistently exhibits $\Delta\approx38$~{meV} for a $T_c=94$~K, as reported by Miyakawa et al.~\cite{Miyakawa99} The~bulk $T_c$ of Tl-2212 films and crystals is typically above 100~K.  Tunneling and Andreev reflection experiments on $c$-axis oriented {Tl-2212} thin films \cite{Giubileo01,Giubileo02} published more recently displayed gap sizes in the same range as in earlier reports, i.e. $20$ and $25$ {meV}. The early tunneling data can be characterized as showing smaller $\Delta$ than expected, large zero-bias conductance values (even above 50\% of the ``background'', i.e.~the~zero-bias normal state conductance extrapolated from high-bias voltage dependence of junction conductance) and poorly defined quasiparticle peaks. In contrast to the tunneling measurements, Raman scattering measurements on {Tl-2212} single crystals by {Kang} \emph{et al.} \cite{Kang97} gave $\mathrm{\Delta_{max} \sim 45}$ {meV}. A~more recent publication of {Wang} \emph{et al.} \cite{Wang03} supported this result with data obtained by infrared spectroscopy of $c$-axis oriented thin film {Tl-2212} samples. The published value was $\mathrm{\Delta_{max}\sim 43}$ {meV}. Both groups considered the~$d$-wave character of superconductivity in Tl-2212 to determine the gap size $\mathrm{\Delta_{max}}$. A summary of all the values is given in {Table} \ref{tab:paramsum}.

\begin{table*}
\caption{\label{tab:paramsum}Overview of superconducting gap size and coupling strength values published previously, measured on thallium cuprates {Tl-2201, Tl-2212, Tl-2223} and on bismuth cuprate {Bi-2212}.}

  \begin{ruledtabular}
    \begin{tabular}{lcccccc}
        Material & Reference & \parbox[c][30pt][c]{1in}{\centering{sample character}} & \parbox[c][30pt][c]{1in}{\centering{experimental method\footnotemark[1]}} & $\mathrm{T_c (K)}$ & $\mathrm{\Delta (meV)}$ & $\mathrm{\frac{ 2\Delta }{ k_BT_c }}$ \\ \hline
            Tl-2201 & \onlinecite{Ozyuzer98} & single crystal & PCT & 91 & $20-22$ & $5.1-5.6$            \\ 
                         & \onlinecite{Ozyuzer99a} & single crystal & PCT & 86 & $19-25$ & $5.1-6.7$          \\ 
                         & \onlinecite{Ozyuzer99b} & single crystal & PCT & 93 & 20 & 5.0                                 \\ 
                         & \onlinecite{Nemetschek93} & single crystal & Raman & 80 & $28-31$ & $8.0-9.0$       \\ 
                         & \onlinecite{Blumberg94} & \parbox[c][30pt][c]{1in}{\centering{single crystal overdoped}} & Raman & 85 & 22 & 6.0             \\ 
                         & \onlinecite{Kendziora96} & \parbox[c][30pt][c]{1in}{\centering{single crystal doping varied}} & Raman & 78 & 24 & 7.2       \\ 
                         & \onlinecite{Gasparov96} & single crystal & Raman & 90 & $27-30$ & $7.0-7.8$        \\ 
                         & \onlinecite{Kang96} & single crystal & Raman & 85 & 29 & 8.0        \\ 
                         & \onlinecite{Gasparov98} & \parbox[c][30pt][c]{1in}{\centering{single crystal doping varied}} & Raman & 80 & 27 & 7.7           \\ \hline
            Tl-2212 & \onlinecite{Huang89}    & single crystal & PCT        & 112   & $16-28$ & $3.3-5.8$ \\ 
                         & \onlinecite{Takeuchi89} & polycrystal    & tunneling & 94.5 & 25           & 6.1           \\ 
                         & \onlinecite{Giubileo01}  & thin film & \parbox[c][30pt][c]{1in}{\centering{break junction tunneling}} & 91 & 25 & 6.3        \\ 
                         & \onlinecite{Giubileo02}  & thin film & \parbox[c][30pt][c]{1in}{\centering{Andreev reflections}} & 104 & 20 & 4.5      \\ 
                         & \onlinecite{Kang97}       & single crystal &Raman  & 102 & 45 & 10.2                    \\ 
                         & \onlinecite{Wang03}      & thin film & IR response & 108 & 43 & 9.2          \\ \hline
            Tl-2223 & \onlinecite{Takeuchi89} & polycrystal & PCT & 114 & $25-35$ & $5.1-7.1$     \\ 
                         & \onlinecite{Vieira90} & polycrystal & STS & --- & $20-24$ & ---       \\ 
                         & \onlinecite{Hoffmann94} & single crystal & Raman & 118 & 38 & 7.4            \\ 
                         & \onlinecite{Mukherjee95} & thin film & Raman & 111 & 33 & 7.0            \\ \hline
            Bi-2212 & \onlinecite{Miyakawa99} & \parbox[c][30pt][c]{1in}{\centering{single crystal doping varied}} & PCT &  95 & $30-40$ & $7.3-9.8$        \\ 
                         & \onlinecite{Ozyuzer00} & \parbox[c][30pt][c]{1in}{\centering{single crystal optimal doping}} & PCT & 95 & 38 & 9.3       \\ 
                         & \onlinecite{Kendziora95} & \parbox[c][30pt][c]{1in}{\centering{single crystal doping varied}} & Raman & 90 & 38 & 8.5      \\ 
                         & \onlinecite{Klein00} & \parbox[c][30pt][c]{1in}{\centering{single crystal doping varied}} & Raman & 95 & 34 & 8.3       \\ 
                         
    \end{tabular}
  \end{ruledtabular}
  \footnotetext[1]{\emph{Abbreviations:} Raman = Raman scattering, PCT = point contact tunneling, STS = scanning tunneling spectroscopy.}
\end{table*}

There is a large difference between Tl-2212 values obtained by bulk probes ({Raman} scattering, {IR} spectroscopy) and the surface probes (electron tunneling, Andreev reflection spectroscopy), which cannot be satisfactorily explained by the difference in the model used in data analysis. The possibility that the observed difference may be inherent in thallium-based superconductors owing to a significantly stronger chemical bonding between layers than {Bi-2212} could be supported by the single-layered {Tl-2201} data \cite{Ozyuzer98,Ozyuzer99b,Nemetschek93,Kendziora96}, where a similar inconsistency is observed. The suggestion fails, though, in case of the tri-layered {Tl-2223} \cite{Takeuchi89,Vieira90,Hoffmann94,Mukherjee95}, where results of surface and bulk probes are closer. Values published in references have been added to Table \ref{tab:paramsum}.

Research of thallium-based cuprates has suffered from the poisonous quality of one of its base chemicals, the~$\mathrm{Tl_2O_3}$. Owing to this, larger part of the world-wide sample manufacturing efforts were focused on compounds safer to work with. Having reviewed the published electron tunneling data we believe that samples available for the experiments might have been of an insufficient crystalline quality (i.e. not single-phase crystals), not uniform oxygen doping or a combination of these, due to difficulties in the fabrication processes. Advan\-ces in the development of manufacturing procedures of thallium cuprate samples \cite{Yan93,Yan94,Ma05} allow us to do tunneling measurements with the currently available high-quality Tl-2212 samples. Results of our point contact tunneling experiments on one {Tl-2212} single crystal and two $c$-axis oriented thin-films are presented below.

The key results are the following. Junctions exhibit improved gap region characteristics including relatively low zero bias conductances and well-defined conductance peaks. The sub-gap region is consistent with a~\mbox{$d$-wave} density of states. A majority of junctions display \mbox{$\Delta>30$~meV}, higher than previous reports, and more consistent with bulk measurements. Finally, there is reproducible evidence of a spectral feature for \mbox{$eV > \Delta$} similar to the~well known dip feature observed~\cite{Romano06} in \mbox{Bi-2212}.

\section{Samples and instrumentation}\label{sec:sampinstr}

Single crystal samples were grown from a stoichiometric mixture of $\mathrm{Ba_2CaCu_2O_x}$ precursor and $\mathrm{Tl_2O_3}$. The~precursor powder was mixed with $\mathrm{Tl_2O_3}$ powder and pressed into a pellet. The pellet was placed in an alumina crucible and covered with $\mathrm{Al_2O_3}$ lids to prevent the thallium vapors from escaping. The mixture was allowed to melt at high temperature in a furnace with steady $O_2$ flow and then cooled down gradually. This procedure is described in detail in Ref.~\onlinecite{Ma05}. The resulting critical temperature of the single crystal was obtained as the onset temperature in bulk $\mathrm{\rho(T)}$ measurement, $\mathrm{T_c}$ = 103 K.

The epitaxial $c$-axis oriented thin films of thickness $\mathrm{\sim400}$ nm were prepared using {D.C.} magnetron sputtering from a stoichiometric target on a $\mathrm{LaAlO_3}$ substrate, and post-deposition annealing. Details of the thin film growth are published elsewhere \cite{Yan93,Yan94}. The thin film samples (TFa and TFb) investigated had $\mathrm{T_c}$ = 106 K and 102 K, respectively, obtained by {SQUID} magnetometer.

Our point-contact tunneling system allows measurements of junctions within a large range (across $\sim 8$ orders of magnitude) of normal-state junction resistances. A differential micrometer is used for fine approach of the tip and to create the point contact junctions. The~junction areas are typically less than, or in the order of, 1~$\mu$m$^2$. Two electrical leads are connected to the~sample using silver paint. Together with the tip they constitute a three-point measurement setup. Sample is affixed to a~holder so that the gold tip approaches along the~\mbox{$c$-axis} of the~sample. Tips are mechanically sharpened and chemically cleaned (aqua regia etching) before every experimental run. Details of the point-contact tunneling measurement system were published in Ref.~\onlinecite{Ozyuzer98a}.

All data presented and discussed in this article were measured using gold tips on samples at 4.2 K. To avoid interference of thermal contraction, the system was cooled down to LHe temperature with tip retracted. Junctions were created and measured after system reached thermal equilibrium. As the tip is moved towards the~sample, the~$I(V)$ signal is monitored on an oscilloscope screen. The~approach is stopped, when a characteristics of a~superconducting tunnel junction appears on the oscilloscope screen. The point contact is open between every two junctions measured to inadvertent data duplicity. A~\mbox{lock-in} technique is used for direct measurement of the~differential conductance, $G(V) = dI/dV$. Both $I(V)$ and lock-in signals are recorded simultaneously in data files for later data processing.

The cuprates are known to have many stable phases and doping levels, and the superconducting parameters have been shown \cite{McElroy05} to vary along the cuprate surfaces at the nanometer length scale, even for very high quality samples. Local surface probes, such as the STM and PCT, are much more sensitive spread of quasiparticle gap sizes much larger than global probes. It has been stated earlier \cite{Ozyuzer99} that the PCT method can produce a variety of junctions on a single sample. In {PCT} the tip can be used to scrape, clean, and in some cases cleave the surface of a sample. Cleaving of a superconductor may result in an attachment of superconducting particles to the metallic tip and creation of a homogeneous superconductor-insulator-superconductor (SIS) junction (see Ref.~\onlinecite{Ozyuzer00}, for example). Care needs to be taken to distinguish the types of junctions in order to interpret the~superconducting gap features correctly. Additionally, the~lack of control over the local properties of the sample surface and the junction atomic-level geometry leads to a varying quality of junctions. Association of many obtained characteristics with a simple enough model is then difficult, or worse, impossible. Therefore, data presented here are representatives of junction characteristics that exhibited superconducting gap features, which is not the~whole set of characteristics obtained on our samples.

\section{Experimental results}

\begin{figure}
  \begin{flushleft}
    \hspace{0in}a) \hspace{0.15in}thin films \hspace{0.81in} b) \hspace{0.15in}single crystal \\
    \includegraphics*[width=3.4in, bb=20 142 595 658]{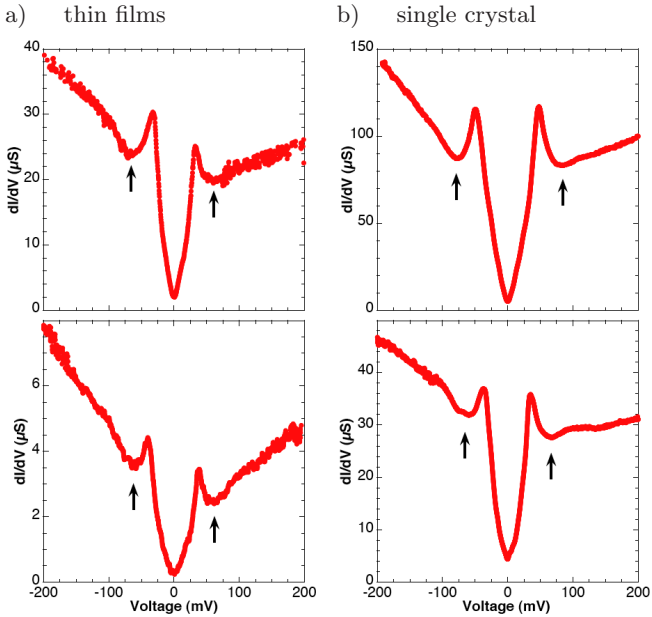}
    \end{flushleft}
  \caption{(Color online) Tunneling junctions data measured on {Tl-2212} a)~thin films and b)~single crystal represented by red (grey) markers. Black arrows have been added at voltages of local minima corresponding to the spectral ``dip'' features. Horizontal scales of the top and bottom plots are the same.} \label{fig:juncdI}
\end{figure}

Figure \ref{fig:juncdI} shows two differential conductance curves for each sample type, thin film and single crystal. All presented data was measured at 4.2 K using a~lock-in amplifier and subsequent calibration with the numerical derivative of current-voltage characteristics. The~data presented here constitutes a substantial improvement from previously published tunneling data on {Tl-2212} in all aspects. The~quasiparticle peaks are very well defined, the zero-bias conductance is low, and the sub-gap regions of the~conductance curves are in an excellent agreement with the $d$-wave model. Additionally, there is evidence of the same ``dip'' feature (indicated by arrows in Fig.~\ref{fig:juncdI}) consistently observed in published {Bi-2212} tunneling data \cite{Ozyuzer99}, which is evidence \cite{Zasadzinski06} of strong electron coupling to a boson mode. We observe that the tunneling background has a strong asymmetric linear dependence $\propto|V|$, which is very reproducible and common to all presented data sets. As discussed in Ref.~\onlinecite{Zasadzinski03}, this linear dependence of tunneling background is likely due to an~additional, inelastic tunneling channel in the junctions. This behavior is common for superconducting cuprates without an easy cleavage plane.

\begin{figure}
  \begin{flushleft}
    \hspace{0in}a) \hspace{0.15in}thin films \hspace{0.81in} b) \hspace{0.15in}single crystal \\
     \includegraphics*[width=3.4in, bb=20 144 595 658]{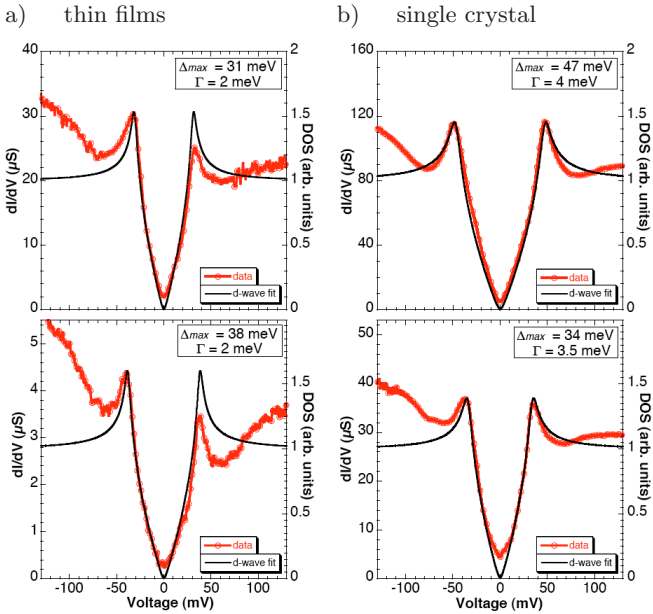}
    
  \end{flushleft}
  \caption{(Color online) Tunneling junctions data represented by red (grey) line with empty circle markers, as measured on {Tl-2212} a)~thin films and b)~single crystal. The~data is here overlaid with the \mbox{$d$-wave} model superconducting density of states, which is represented by black line with small dot markers. Parameters $\Delta_{max}$ and $\Gamma$ of the~\mbox{$d$-wave} model curves are displayed in text boxes in upper right corners of the respective plot areas. Horizontal scales of the~top and bottom plots are the same.} \label{fig:juncnorm}
\end{figure}

Because of the strong linear character of the background, direct normalization of the differential conductance curves is not sensible. This is because the additional inelastic channel does not to conserve states. Differential conductance curves from Figure \ref{fig:juncdI} were fit, without normalization, with $d$-wave tunneling DOS model, and the model curves were scaled together with the data for display in Figure \ref{fig:juncnorm}. Thermal smearing present at $T= 4.2$ K and inelastic scattering rate $\Gamma$ were included in the superconductor-insulator-normal metal (SIN) tunneling model. Precision of this fit is not a serious issue in our context, because the $\Delta_{max}$ values extracted from the fit differ only by $\sim1$ meV in the range of acceptable values of $\Gamma$. Although there is a spread of $\Delta_{max}$ values, they are consistently larger than those derived from previously published ones and in agreement with results of optical experiments. 

\begin{figure}
  \begin{flushleft}
    \hspace{0in}a) \hspace{0.15in}thin films \hspace{0.81in} b) \hspace{0.15in}single crystal \\
     \includegraphics*[width=3.4in, bb=17 253 593 540]{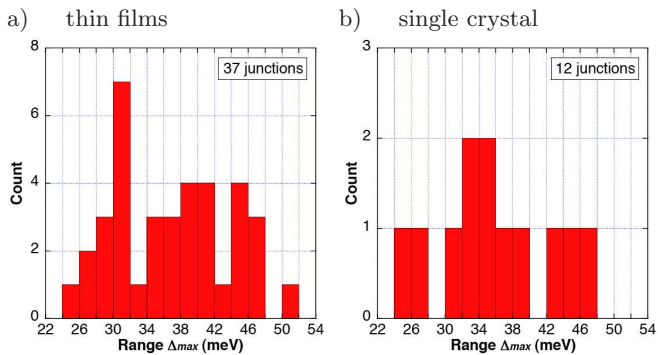}

  \end{flushleft}
  \caption{(Color online) Histograms of $\Delta_{max}$ values of SIN junctions measured on Tl-2212. a)~37 SIN junctions on 2 thin film samples, and b)~12 junctions on 1 single crystal sample. Note that 70\% junctions overall exhibited $\Delta_{max} \geq 30$ meV.} \label{fig:hist}
\end{figure}

\begin{figure}
  \begin{center}
    \includegraphics*[width=3.4in, bb=20 160 590 620]{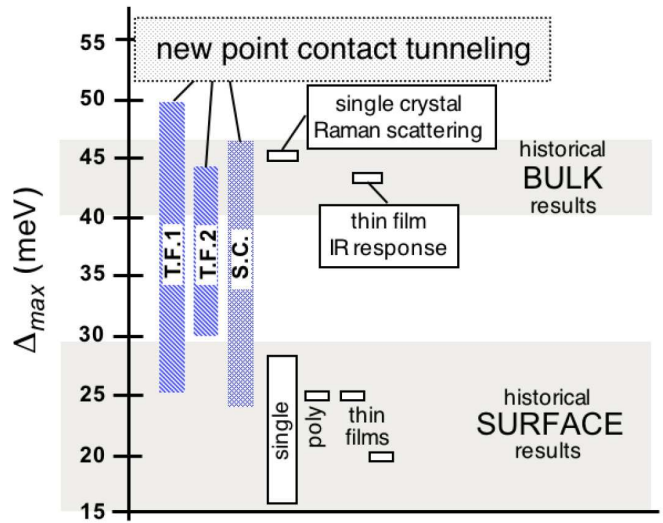}

  \end{center}
  \caption{(Color online) Summary of new $\Delta_{max}$ measured on \mbox{Tl-2212}. Graphics displays new tunneling results introduced in this article using the blue (grey) pattern-filled bars. The~range representing thin film no.1 (T.F.1) is based on 29 junctions, range of thin film no.2 (T.F.2) is based on 8~junctions and the bar representing the~single crystal (S.C.) is based on 12 junctions, all of which were included in histograms in Figure \ref{fig:hist}. For review, historical {Tl-2212} results summarized in Table \ref{tab:paramsum} were added and they are represented by blank outlined boxes.} \label{fig:sumbox}
\end{figure}

The same fitting as shown in Figure \ref{fig:juncnorm} was applied to all measured $dI/dV$ curves, which featured sharp quasiparticle peaks. There were 37 such junctions on thin film samples and 12 on the single crystal. Thus obtained $\Delta_{max}$ values are presented in histograms in Figure \ref{fig:hist}. The histograms show that $\Delta_{max}$ values consistently fall in ranges $\approx 25 - 50$ for thin films and $\approx 24 - 47$ for single crystals. More than 70\% of the junctions exhibited $\Delta > 30$~meV. Variation of $\Delta$ throughout the measured set of tunneling junctions can be due to several reasons, intrinsic or extrinsic. In case of the intrinsic reasons we can point out the measurements of McElroy \emph{et~al.}~\cite{McElroy05}, who reported large doping variation along Bi-2212 cuprate surface. For the~extrinsic causes of the~spread of values we can mention the possibility of surface damage-induced change in oxygen doping. The spread in $\Delta_{max}$ not inconsistent with historical results of optical methods and there was little correlation between measurement conditions and the data taken. On one thin film sample, there seemed to be the tendency to measure smaller gaps at the~beginning of the experiment and larger towards the~end, suggesting the repeated contact of the tip was scraping through the surface. But this was not the case with other two {Tl-2212} samples. This observation indicates that the influence of extrinsic doping changes is minor. We conclude that the spread of gap sizes is largely intrinsic to the samples. These ranges have been put together in the graphics in Figure \ref{fig:sumbox} for comparison with the~historical Tl-2212 values, which we already summarized in Table \ref{tab:paramsum} in the {Introduction}. 

To interpret the tunneling data correctly and to obtain the true $\Delta_{max}$, we had to make sure that the junctions included in the analysis had no other but SIN character. The alternative junction type, which has the same general features, is the SIS break junction. An incorrectly applied SIN model would result in $\sim 2\times$ the actual $\Delta_{max}$. Observing that the historical inconsistency in the $\Delta_{max}$ values determined by bulk and surface probes is almost by a factor of 2, we consider this point worthy of special attention.

General features common to SIN and SIS tunneling characteristics of high-$\mathrm{T_c}$ superconductors are existence of quasiparticle peaks, approximately cusp-like shape of the sub-gap region (in agreement with $d$-wave model, see Ref.~\onlinecite{Won94}) and an additional high-bias feature at voltages above $\Delta_{max}/e$, the so called ``dip feature'' \cite{Romano06,Zasadzinski06}. One effect, which in many cases labels the SIS junctions, is the~zero-bias conductance peak due to {Josephson} tunneling. Josephson conductance peak can be suppressed, however, in high-resistance SIS junctions. At the same time, small zero-bias peaks, which are due to effects other than superconductivity, can occur in SIN junctions as well. Normal state resistances (zero bias background estimate) of tunneling junctions we measured on both sample types spanned approximately from 500 $\Omega$ to 2 M$\Omega$. Junctions with a distinct zero-bias peak were eliminated from the analysis early on. Very low (less than 5\% of peak height) zero-bias peaks were observed in some other junctions, but this occurrence was uncorrelated with junction resistance or peak position. Therefore, we could not identify those junctions directly as SIS. Other features of SIN and SIS tunneling conductance curves had to be looked upon to distinguish between them. Comparison with published point-contact tunneling data on Bi-2212 was very helpful.

The widely recognized features distinguishing between SIN and SIS $d$-wave model characteristics are the detailed shape of sub-gap region, and higher (peak-to-background ratio) quasiparticle peaks of the SIS characteristics. There are other distinctive properties of experimental $dI/dV$ characteristics, which cannot be derived from the $s$-wave or $d$-wave models. Homogeneous SIS junctions have symmetrical characteristics, both in peak height and background shape. The strong asymmetry of tunneling background we observed is not consistent with SIS junction character, but SIN characteristics are often asymmetrical \cite{Ozyuzer98}. Another distinguishing element is the position and magnitude of the ``dip feature'', which has been consistently observed in Bi-2212 and is likely \cite{Zasadzinski06} a result of strong electron-boson coupling in HTS. The dips are less pronounced in SIN curves, while positioned at $V^{dip}_{SIN}\approx 2\Delta_{max}/e$. In SIS characteristics dips are symmetrical and sharper (these dips were observed to go negative in some {SIS} junctions on \mbox{Bi-2212}) and it is located at $V^{dip}_{SIS}\approx 3\Delta_{max}/e$ (see Ref.~\onlinecite{Zasadzinski06}). The set of $dI/dV$ curves displayed in Figures \ref{fig:juncdI} and \ref{fig:juncnorm} show that all conductance curves are asymmetrical to some degree. At the same time, the ``dip'' features accented in Figure \ref{fig:juncdI} are present and located consistently at $V^{dip}\approx 1.8 - 2.3 \Delta_{max}/e$. Finally, the sub-gap regions are consistently more alike the SIN model characteristics, rather than the SIS ones (for SIS model curves see for example Ref.~\onlinecite{Romano06}). Using these attributes we were able to justify that all characteristics included in our analysis were of SIN type, and that superconducting gap parameters summarized in Figures \ref{fig:hist} and \ref{fig:sumbox} are correct. 

\section{Conclusions}

We have created and measured 49 high-quality SIN tunneling junctions on three different Tl-2212 cuprate samples, which exhibited $\Delta_{max}$ in the range $25-50$~meV for $c$-axis oriented thin films and $24-47$ meV for a single crystal. About 80\% of all the $\Delta_{max}$ values were greater than 30 meV, which is in stark contrast with previous tunneling studies of {Tl-2212} samples \cite{Huang89,Takeuchi89}, where published $\Delta_{max}$ were all below 30 meV, most of them in the range 20--25 meV. These results are most likely a consequence of improved sample quality and homogeneity. The~new $\Delta_{max}$ values were obtained from tunneling characteristics with sharp quasiparticle peaks, low zero-bias conductance, and an obviously $d$-wave character of the sub-gap region. There is a clear consistency between data measured on thin films and single crystals. Our results, including the spread of energy gaps $\Delta_{max}$ are in agreement with bulk measurements on Tl-2212 thin films and single crystals published in \cite{Kang96} and \cite{Wang03}. As there was no convincing correlation between the gaps sizes measured and experiment progress, we conclude that the spread of values is most likely due to a~real spread of doping across the~samples, while the influence of the measurement procedure on our result is minute. Additionally, the tunneling characteristics exhibit a high-bias "dip" feature, which has been consistently observed in {Bi-2212} and is likely a result of strong coupling of the superconducting electrons to a boson. It has been proposed that the dip in tunneling is linked to peak in optical self-energy~\cite{Zasadzinski06} in Bi-2212. Now a similar linkage is formed in Tl-2212, which also exhibits~\cite{Wang03} a peak in optical self-energy.

\begin{acknowledgments}
The first author would like to thank Dr.~L\"{u}tfi \"{O}zy\"{u}zer for valuable advice and leadership. Dr.~N.~L.~Wang acknowledges support from the~National Science Foundation of China and the Ministry of Science and Technology of China (973 project No.~2006CB601002). The submitted manuscript has been authored by the UChicago Argonne, LLC, Operator of Argonne National Laboratory under the Contract No. DE-AC02-06CH11357 with the U.S. Department of Energy. 
\end{acknowledgments}

\bibliographystyle{apsrev}

\end{document}